\documentclass[sigconf, review,anonymous=false]{acmart}

\AtBeginDocument{%
 \providecommand\BibTeX{{%
   \normalfont B\kern-0.5em{\scshape i\kern-0.25em b}\kern-0.8em\TeX}}}

\setcopyright{none}
\renewcommand\footnotetextcopyrightpermission[1]{}
\usepackage{amsmath,amsfonts}
\usepackage{algorithmic}
\usepackage{comment}
\usepackage{graphicx}
\usepackage{textcomp}
\usepackage{xcolor}
\usepackage{subfiles}
\usepackage{xurl}
\usepackage{array}
\usepackage{multirow}
\usepackage{xspace}
\usepackage{enumitem}
\usepackage{hyperref}
\usepackage{footmisc}
\usepackage{tabularx} 
\begin{document}

\title{Security-First Approach to API Pipeline Development with Zero-Trust Architecture}

\author{Mahima Agarwal}
\email{mahima.agarwal14@gmail.com}
\affiliation{%
 \institution{Microsoft Corporation}
  \city{Vancouver}
  \country{Canada}
}

\author{Keshav Ranjan}
\email{ranjan.keshav11@gmail.com}
\affiliation{%
 \institution{Palo Alto Networks}
  \city{Santa Clara}
  \country{United States}
}

\begin{abstract}

Modern enterprises face an accelerating onslaught of API-targeted threats amid a rapidly expanding attack surface. Record volumes of software vulnerabilities continue to accelerate dramatically, with 28,818 CVEs disclosed in 2023 (a 38\% jump from 2022) and 40,009 CVEs in 2024 (another 38\% increase)\cite{gamblin2025cve}, while the average time-to-exploit (TTE) of new flaws shrank to mere days (approximately 5 days in 2023, down from 32 days in 2021)\cite{zorz2024tte}. At the same time, API usage dominates web traffic and has become a primary vector for breaches - 99\% of organizations experienced API security incidents in the last year, with 22\% suffering actual data breaches via APIs (based on industry vendor research)\cite{salt2025api}. This paper proposes a comprehensive "security-first" framework for API pipeline development, leveraging Zero-Trust Architecture principles within DevSecOps practices to counter these trends. We introduce a five-pillar approach encompassing Governance \& Planning, Secure Design, Continuous Testing, Pipeline Controls, and Runtime Protection, aligned with industry standards (OWASP API Security Top 10 2023, NIST Secure Software Development Framework) and recent cybersecurity advisories. The results show significant improvements in vulnerability mitigation and breach prevention (e.g., 30\% reduction in security incidents and 40\% fewer post-release vulnerabilities in representative case studies), highlighting the positive impact of proactive security integration. The paper concludes with a discussion on implementation challenges, the evolving threat landscape, and recommendations for organizations to adopt a security-first pipeline with Zero-Trust to fortify API development against current and future threats.
\end{abstract}

\keywords{API Security, DevSecOps, Business Logic Vulnerabilities, Runtime API Protection, Threat Modeling}
\maketitle
\pagestyle{plain}

\section{Introduction}
APIs have become the backbone of modern applications and services, but their proliferation has introduced significant security challenges. Recent industry studies and breach reports reveal a concerning rise in API-related security incidents. A 2025 survey found that 99\% of organizations encountered API security problems over the past year, and over one-fifth suffered an API-linked data breach\cite{salt2025api}. Note: Some statistics in this paper are derived from vendor-sponsored research and should be interpreted with awareness of potential commercial bias.

\subsection{Vulnerability Trends}
The vulnerability landscape has experienced unprecedented acceleration in recent years. CVE disclosure rates have shown consistent 38\% year-over-year growth: from approximately 21,000 CVEs in 2022, to 28,818 in 2023, to 40,009 in 2024\cite{gamblin2025cve}. This represents an average of 108 new vulnerabilities disclosed daily throughout 2024, with May 2024 recording the highest monthly total of 5,010 CVEs.

This dramatic increase reflects not only growing software complexity and broader technology adoption but also improved vulnerability discovery and disclosure processes. However, it also means that over 15\% of all CVEs ever published occurred in 2024 alone, creating unprecedented challenges for vulnerability management teams.

Attackers increasingly target APIs for sensitive data and business logic, often using "low and slow" techniques to evade detection\cite{salt2024owasp}. These trends coincide with a broader surge in software vulnerabilities and faster exploitation. The year 2023 saw 28,818 disclosed CVEs (up 38\% from 2022), followed by 40,009 CVEs in 2024 (another 38\% increase) \cite{gamblin2025cve} and threat actors now exploit published flaws within days - the average time from disclosure to first exploit has plummeted to \textasciitilde{}5 days\cite{zorz2024tte}.

Notably, 70\% of top exploited bugs in 2023 were zero-day vulnerabilities, indicating that attackers are pouncing on unknown flaws and outpacing patch cycles\cite{zorz2024tte}. In some cases, up to 25\% of vulnerabilities are exploited on the same day they are disclosed\cite{firecompass2025tte}, leaving organizations with a vanishingly small window to respond.

\subsection{API Attack Surface Growth}
Compounding the risk, APIs dramatically expand the attack surface: they now constitute roughly 71\% of all web traffic\cite{imperva2024api} and often expose numerous endpoints beyond traditional web interfaces. The OWASP API Security Top 10 (2023) highlights how API-specific issues like broken object level authorization (BOLA), broken user authentication, excessive data exposure, lack of resource limiting, and other flaws are prevalent and actively exploited\cite{owasp2023api}.

For instance, BOLA (API1:2023) remains the most common API vulnerability - it's responsible for \textasciitilde{}40\% of API attacks by allowing unauthorized access to data using object identifiers\cite{salt2024owasp}. Despite such well-known risks, many organizations lag in addressing them: an estimated 80\% of API attack attempts leverage weaknesses from the OWASP API Top 10, yet only about 53\% of organizations focus on these API-specific vulnerability categories\cite{salt2025api}.

\subsection{The Need for Paradigm Shift}
This gap underscores the need for stronger governance and proactive security measures around API development. Considering this threat landscape, a paradigm shift is required in how APIs are built and secured. Traditional perimeter-based defenses and after-the-fact security testing are proving insufficient against the current scale and speed of vulnerability disclosure and exploitation.

Instead, organizations are moving toward a security-first approach that embeds robust security practices throughout the API lifecycle - from initial design to deployment and production. Equally important is adopting a Zero-Trust Architecture stance: assume no implicit trust, continuously verify every entity and interaction, enforce least privilege, and operate under the expectation of breach.

Zero-Trust principles, often summarized as "verify explicitly, use least-privilege access, and assume breach"\cite{unosecur2025nistzt}, align closely with DevSecOps ideals of integrating security into every stage of software delivery. By operationalizing these principles within CI/CD pipelines, development teams can significantly harden their API ecosystems against modern attacks.

\subsection{Paper Contribution}
This paper presents a structured framework for achieving a security-first API pipeline under a Zero-Trust model. We outline a five-pillar approach - Governance \& Planning, Secure Design, Continuous Testing, Pipeline Controls, and Runtime Protection - that together provide end-to-end safeguards.

In the sections that follow, we review related work and standards (OWASP, NIST SSDF, etc.), describe our high-level solution, and then detail the methodology for each pillar. We show how techniques like threat modeling, secure coding standards, automated security testing (SAST/DAST/IAST), supply chain security controls, and runtime monitoring can be holistically integrated.

Real-world case studies and metrics are discussed to illustrate the benefits of this approach, such as faster incident response, reduced vulnerabilities, and improved resilience. Ultimately, by building security into the API pipeline and embracing Zero-Trust principles, organizations can substantially reduce their risk of API breaches while maintaining the agility of DevOps.

\section{Literature Review}

\subsection{Recent API Threat Landscape}
Recent literature underscores that API vulnerabilities and attacks have become a top enterprise security concern. The OWASP API Security Top 10 (2023 edition) distills the most critical API risks, reflecting lessons learned since the previous 2019 list\cite{owasp2023api}. The updated 2023 list identifies issues such as:

\begin{itemize}
\item API1: Broken Object Level Authorization (BOLA) - the leading vulnerability, enabling attackers to manipulate object identifiers to access data without proper authorization\cite{salt2024owasp}
\item API2: Broken Authentication - flaws in token handling or credential management that allow account compromise\cite{salt2024owasp}
\item API3: Broken Object Property Level Authorization - a newly defined category merging excessive data exposure and mass assignment issues, where lack of granular property-level checks leads to unintended data access\cite{owasp2023api,salt2024owasp}
\item API4: Unrestricted Resource Consumption - failure to enforce rate limiting or quotas, risking denial-of-service through excessive usage\cite{owasp2023api}
\item Other issues like function-level authorization gaps, server-side request forgery (SSRF), security misconfigurations, improper asset (inventory) management, and unsafe consumption of third-party APIs\cite{owasp2023api}
\end{itemize}
These documented risks are not merely theoretical - many have been observed in high-profile incidents. For example, broken authorization vulnerabilities (BOLA/BFLA) have been implicated in numerous data breaches where one customer could fetch another's records by modifying an ID, and excessive data exposure in APIs has led to leakage of sensitive personal data when developers returned more information than intended.

The State of API Security reports by industry groups provide empirical confirmation: according to Salt Security's 2025 report (noting this is vendor-sponsored research), nearly all organizations (99\%) experienced API security issues, and 22\% had an API-related breach in the past year\cite{salt2025api}. The same report highlights that attackers are increasingly probing API business logic with subtle, slow attacks to avoid detection\cite{salt2024owasp}. These findings reinforce why dedicated API security practices - beyond general web security - are crucial.

\subsection{Secure Frameworks}
To combat rising vulnerabilities, the industry has coalesced around integrating security into the software development lifecycle (SDLC), often termed DevSecOps. One cornerstone is the NIST Secure Software Development Framework (SSDF), which provides a set of fundamental practices for "secure by design" development.

NIST SSDF (SP 800-218 v1.1) groups recommended practices into four primary categories\cite{nist2022ssdf}:

\begin{itemize}
\item Prepare the Organization (PO) - ensure people, processes, and tooling are in place for security (e.g. training developers, establishing security requirements and governance)
\item Protect the Software (PS) - safeguard code, dependencies, and environments from tampering and unauthorized access (for instance, controlling access to repositories, using signing for code and builds)
\item Produce Well-Secured Software (PW) - implement activities during development to minimize vulnerabilities in releases (such as threat modeling, secure coding standards, static and dynamic testing, dependency scanning)
\item Respond to Vulnerabilities (RV) - plan for and address residual vulnerabilities (e.g. vulnerability disclosure programs, patch response processes)
\end{itemize}
Following such practices has been shown to reduce the density of security flaws and the impact of exploits\cite{nist2022ssdf}. Government and industry guidelines now strongly advocate for these secure development practices. In fact, a joint cybersecurity advisory by CISA and allied agencies explicitly recommends software producers to "implement secure-by-design practices into each stage of the SDLC," following frameworks like NIST SSDF\cite{cisa2024exploited}.

The advisory also highlights specific tactics like conducting threat modeling throughout development and improving testing rigor to catch issues earlier\cite{cisa2024exploited}. Another important aspect is software supply chain security: with the rise of attacks on build pipelines and open-source components, measures such as Software Bills of Materials (SBOMs), dependency vulnerability scanning (SCA tools), and integrity verification (e.g. Sigstore for artifact signing) are increasingly mandated in standards and regulations\cite{filippidis2023cicd}.

These preventative controls align with the "Shift Left" philosophy in DevSecOps - addressing security in the early phases (planning, coding, build) so that fewer issues reach production.

\subsection{Zero-Trust Architecture (ZTA)}
Zero-Trust has emerged as a guiding security model in response to perimeter breaches and lateral attacker movement. Traditional network-based defenses are insufficient when attackers can easily bypass perimeter controls (through stolen credentials, supply chain compromises, etc.).

Zero-Trust Architecture, as defined by NIST SP 800-207, operates on golden rule "never trust, always verify." All access requests - whether from external or internal sources - must be explicitly authenticated and authorized based on context, with fine-grained least-privilege policies, and activities are monitored under the assumption of an eventual breach\cite{unosecur2025nistzt,riskrecon2023nistzt}.

In practical terms, ZTA shifts security to focus on identity, devices, and workloads rather than trusting network location. Key tenets include:

\begin{itemize}
\item Strong continuous identity verification (e.g. MFA everywhere, device identity checks)
\item Dynamic access control (granting minimal rights just-in-time)
\item Micro-segmentation of networks and services
\item Pervasive monitoring and analytics for anomalous behavior\cite{unosecur2025nistzt}
\end{itemize}
Microsoft and others distill Zero-Trust into three core principles:

\begin{itemize}
\item Verify Explicitly (authenticate and authorize every action, no implicit trust based on network or credentials)
\item Least Privilege Access (only grant the minimal access required and for the shortest time needed, to limit blast radius)
\item Assume Breach (design as if an attacker is already in the environment - implement segmentation, encryption, and continuous monitoring to contain and respond) \cite{unosecur2025nistzt}
\end{itemize}
These principles have broad applicability, not only for enterprise IT access but also within software delivery pipelines and cloud architectures. Recent works have started exploring the fusion of Zero-Trust with DevSecOps. Filippidis (2023) discusses applying NIST's zero-trust pillars to CI/CD pipeline design, noting that CI/CD systems inherently hold high privileges (to deploy and modify infrastructure) and thus are high-value targets for attackers\cite{filippidis2023cicd}.

Breaches like the CircleCI incident (January 2023) - where an CI service was compromised, exposing numerous organizations' secrets - highlight that compromising a pipeline can directly lead to production system compromise\cite{filippidis2023cicd}. Therefore, principles such as using one-time credentials, verifying code integrity, and segmenting pipeline responsibilities can significantly reduce the risk of CI/CD compromise\cite{filippidis2023cicd}.

In summary, Zero-Trust offers a strategic mindset and set of controls that, when combined with DevSecOps, can greatly harden the software supply chain and runtime environment against both external and insider threats.

\subsection{Related Work}
There is growing evidence that adopting a security-first DevOps approach yields measurable improvements. For example, a Practical DevSecOps case study reported a tech firm that embedded security metrics and checks into its CI/CD pipeline saw a 30\% reduction in security incidents and substantially faster detection/remediation times \cite{practicaldevsecops2025metrics}.

In another case, a financial services company improved cross-team collaboration (developers, ops, security) on threat mitigation, resulting in a 40\% decrease in vulnerabilities detected after release (i.e., fewer issues escaped into production) \cite{practicaldevsecops2025metrics}. Note: These improvement statistics represent specific case studies and may not reflect industry-wide averages. These outcomes echo findings from industry surveys that correlate high DevSecOps maturity with lower security failure rates.

It is worth noting, however, that achieving such results requires cultural and process changes, not just tools. Common challenges noted in studies include resistance to change, the need for upskilling teams in security, and the complexity of integrating multiple security tools into fast-moving pipelines\cite{dragonspearsdevsecops,practicaldevsecops2025metrics}.

Nonetheless, the literature consistently points to the benefit of moving from reactive, perimeter-focused security to a proactive, design-integrated model. Our work builds on these insights by providing a structured framework tailored to API development, which by its nature demands careful handling of authorization, data validation, and monitoring. In the following sections, we describe our proposed solution approach and then delve into each pillar in detail, drawing connections to the discussed standards and prior works.
\section{High-Level Solution Approach}

At a high level, we propose a Five-Pillar Security Framework for API pipeline development that embeds security practices across every phase of the software lifecycle. The five pillars - Governance \& Planning, Secure Design, Continuous Testing, Pipeline Controls, and Runtime Protection - correspond to sequential stages and focus areas in the DevSecOps pipeline.

By addressing security in each pillar, the framework ensures comprehensive coverage from the inception of an API project through its deployment and operation. The approach is "security-first" in that it prioritizes risk management and threat mitigation activities early and continuously, rather than treating security as an afterthought or a final gate.

Each pillar incorporates Zero-Trust principles to eliminate implicit trust and enforce verification at every step. Below is an overview of the pillars and how they interconnect:

\subsection{Governance \& Planning}
This pillar sets the foundation. It involves defining security requirements, policies, and compliance objectives at the outset of API development. Activities include:

\begin{itemize}
\item Establishing secure development standards (aligned with frameworks like NIST SSDF)
\item Performing risk assessments
\item Planning for threat modeling
\item Assigning roles (e.g. security champions in dev teams)
\item Providing training so that everyone involved understands the security expectations
\end{itemize}
Planning with a security-first mindset ensures that subsequent development work is informed by clear security objectives (for example, mandating adherence to OWASP Top 10 guidelines, or planning to implement authentication/authorization using Zero-Trust tenets from day one).

\subsection{Secure Design}
In this stage, architects and developers design the API and system with security built in. Secure design practices involve applying design principles such as least privilege, fail-safe defaults, and privacy by design. A key component here is threat modeling - systematically analyzing the proposed API design for potential threats and abuse cases.

Using methodologies like STRIDE or LINDDUN, teams identify how an attacker might exploit the API's logic, endpoints, or data flows, and then incorporate mitigations into the design\cite{practicaldevsecops2025threatmodel}. For example, if threat modeling reveals a risk of broken object level authorization (API1), the design can specify that every API endpoint will enforce object-level ACL checks.

Design reviews should also cover:

\begin{itemize}
\item Secure authentication flows (avoiding things like reusable tokens or weak API keys)
\item Input validation strategies to prevent injections
\item Proper use of encryption
\end{itemize}
By the end of this phase, security is "baked into" the API architecture and documented as security requirements. Zero-Trust is infused by design - e.g., plan that every request will be authenticated and authorized, no open trusts within microservices, and segmentation of components so a breach in one does not compromise all.

\subsection{Continuous Testing}
This pillar ensures that as code is written and integrated, it is continuously evaluated for security flaws. It corresponds to the CI (Continuous Integration) part of the pipeline where automated security tests are run on every build or release candidate. Key components include:

\begin{itemize}
\item Static Application Security Testing (SAST) - analyzing source code or binaries for vulnerabilities (such as SQL injection, hardcoded secrets, insecure configurations) early in development
\item Dynamic Application Security Testing (DAST) - executing the running API (often in a test environment) and probing it for vulnerabilities (like authentication bypass, XSS, etc.)
\item Software Composition Analysis (SCA) - scanning open-source libraries and dependencies for known CVEs and license risks
\item API contract testing and fuzzing - sending unexpected or random inputs to API endpoints to find edge-case failures
\item Automated authorization testing - ensuring endpoints properly enforce user roles and data scoping
\end{itemize}
Continuous testing means this security checks are integrated into the CI/CD pipeline: whenever a developer commits code, the pipeline triggers SAST scans; when an API build is deployed to a test environment, DAST and integration tests run automatically. The goal is to catch and fix issues before they progress down the pipeline.

This pillar implements the "verify explicitly" principle in an automated fashion - every code change is verified against security policies and test cases. It also supports an "assume breach" mentality by actively testing for possible breach scenarios continually.

\subsection{Pipeline Controls}
Pipeline controls refer to securing the CI/CD pipeline itself and the software supply chain. This pillar is about protecting the build and deployment process from tampering or misuse, echoing the NIST SSDF's "Protect the Software (PS)" group. Fundamental aspects include:

\begin{itemize}
\item Credential and secret management - using ephemeral, short-lived credentials (for instance, employing federated identity tokens for CI jobs rather than long-lived static secrets)\cite{filippidis2023cicd}
\item Integrity checks - requiring signed commits and verifying signatures, signing build artifacts and verifying signatures before deployment
\item Role-based access and environment segmentation - build agents and deployment tools should run with minimal privileges
\item Binary provenance - recording hashes and build metadata to ensure only authorized and untampered artifacts are promoted
\end{itemize}
Recent attacks on CI/CD (such as malicious code injection into build processes, or public CI services being abused to exfiltrate secrets) indicate these controls are not merely theoretical. By applying a Zero-Trust lens, we treat the pipeline as an untrusted environment - each step must authenticate and be authorized to perform actions, and no single breach of one component should automatically grant access to everything.

\subsection{Runtime Protection}
Even after secure development and deployment, APIs face live threats in production. The Runtime Protection pillar focuses on operational defenses and continuous monitoring for APIs in their production environment (the "shift right" part of security). This includes:

\begin{itemize}
\item API gateways or Web Application Firewalls (WAF) with rules tailored to API traffic
\item Runtime Application Self-Protection (RASP) techniques to detect and thwart attacks in real-time
\item Anomaly detection - leveraging logs and telemetry to spot suspicious usage patterns
\item Continuous authentication (e.g., adaptive MFA challenges if a session's behavior is abnormal)
\item Environment and infrastructure monitoring - using IDS/IPS, container security monitoring, and cloud security posture management
\end{itemize}
An important aspect of runtime protection is the feedback loop - information gathered in production should feed back into the development cycle. This pillar enforces Zero-Trust in production: every request is scrutinized, every user's access is minimized, and we operate under the assumption that a breach may occur, so continuous monitoring and response capabilities are in place.

\subsection{ZTA Integration Framework}
\begin{itemize}
\item Zero-Trust is not a separate pillar, but rather an overarching paradigm that influences all pillars:
\item In Governance, we adopt policies of never implicitly trusting any system or actor
\item In Design, we plan for least-privilege and extensive authentication
\item In Testing, we verify everything continuously
\item In Pipeline Controls, we treat the CI/CD infrastructure as hostile unless proven otherwise
\item In Runtime, we relentlessly verify and monitor
\end{itemize}

\begin{figure}[t]
\centering
\includegraphics[width=1\linewidth]
{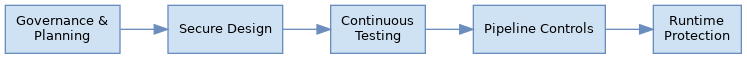}
\caption{Proposed five pillars form a lifecycle framework integrating security from project inception to production runtime.}
\label{fig:five-pillars}
\end{figure}

\section{Detailed Solution and Methodology}
This section will delve into each pillar discussed in previous section, illustrating how to operationalize the concepts with technical specifics and workflow diagram.

\begin{figure*}[!t]
\centering
\includegraphics[width=1\linewidth]
{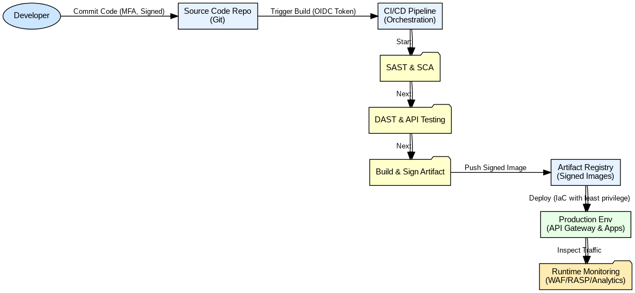}
\caption{Secure API pipeline architecture incorporating Zero-Trust integration points across CI/CD stages.}
\label{fig:secure-api-pipeline}
\end{figure*}

\subsection{Governance \& Planning}
The Governance \& Planning pillar establishes the strategic security foundation before any code is written. In practice, this means instituting organizational policies, standards, and processes that will guide secure API development.

\subsubsection{\bf Security Requirements Framework}
Initially, a security requirements framework should be defined for all API projects. This could draw from industry standards: for example, mandate compliance with the OWASP API Security Top 10 (ensure each project plans mitigations for those top 10 risks) and relevant regulatory requirements (such as GDPR for privacy, PCI DSS if handling

payments, etc.). By setting these requirements upfront, developers and product managers know the security bar they must meet.

\subsubsection{\bf Risk Assessment}
A key governance practice is performing an initial risk assessment during the planning phase of a new API or feature. This involves identifying:

\begin{itemize}
\item Data classification (will the API handle sensitive personal data, financial info, etc.?)
\item Trust levels of clients (public vs internal API)
\item Potential business impact of a breach
\end{itemize}
High-risk APIs (e.g., those exposing critical transactions or sensitive data) would then require more stringent controls and review. The outcome of risk assessment feeds into planning security testing scopes, deciding if a formal architecture risk analysis is needed, etc.

\subsubsection{\bf Threat Modeling Planning}
Another governance element is building a threat modeling plan into the project. Management should require that a threat modeling exercise be completed in the design phase (Secure Design pillar), and allocate the necessary resources/time for it. If the organization has security engineers or a Center of Excellence, they should engage with development teams at this stage.

Roles and responsibilities are defined - for instance, naming a Security Champion in the development team who liaises with the security team and ensures security tasks are on the backlog. This aligns with best practices of cross-functional responsibility: security is not only the security team's job, but everyone's job.

\subsubsection{\bf Compliance and Policy Alignment}
From a compliance perspective, Governance \& Planning includes aligning the pipeline with frameworks like NIST SSDF or ISO 27001. This means documenting processes for code review, testing, and release that incorporate security steps. Some organizations create an internal Secure SDLC policy document that all teams must follow, which outlines stage-by-stage what security activities must happen (for example: "Threat modeling required for all new high-risk features; SAST must run on every commit; all critical vulnerabilities must be remediated before release" etc.).

\subsubsection{\bf Tooling and Training}
Additionally, governance sets up the tooling and training needed for success. Before development starts, the organization should provision necessary security tools (SAST scanners, SCA tools, DAST services, etc.) and make them easily accessible via the CI/CD pipeline. Developers and DevOps personnel should be trained on these tools and on secure coding practices for APIs.

\subsubsection{\bf Incident Response Preparation}
Lastly, planning includes incident response preparation for API security events. Even at project start, one should plan how a security issue will be handled: ensure there is a process to receive vulnerability reports (perhaps through a bug bounty or disclosure program), and that the team knows the steps for patching a critical API bug quickly.

In summary, Governance \& Planning creates the environment and expectations for secure API development. It's an ongoing effort - governance doesn't end after initial planning, but continues through periodic audits, risk reviews, and updates to policies as threats evolve.

\subsection{\bf Secure Design}
Secure Design is the pillar where high-level ideas are translated into concrete, secure architecture decisions for the API. In this stage, architects and senior developers work to ensure that the system's design inherently reduces risk.

\subsubsection{\bf Threat Modeling Workshop}
A crucial activity here is conducting a Threat Modeling workshop for the new API or feature. The team (including developers, architects, and security experts) creates data flow diagrams of the API, enumerates entry points, trust boundaries, and assets, then systematically brainstorms potential threats using frameworks like STRIDE (Spoofing, Tampering, Repudiation, Information Disclosure, Denial of Service, Elevation of Privilege).

For each identified threat, mitigation strategies are devised and fed back into the design. For example, threat modeling an e-commerce API might reveal a threat of mass assignment (automatically binding client inputs to object properties could allow overwriting of fields). The mitigation could be designing the API to whitelist which fields can be updated (to prevent unauthorized property changes) - directly addressing OWASP API3:2019/2023 issues.

\subsubsection{\bf Security Controls and Patterns}
One outcome of threat modeling and secure design is deciding on the security controls and patterns to implement. For instance, an API dealing with sensitive data should be designed to require strong authentication (possibly mutual TLS for client authentication or OAuth 2.0 with short-lived JWT access tokens).

The design would include an authorization scheme, e.g., role-based access control or attribute-based access control, to ensure every request's access rights are evaluated. If B2B or internal, using network-level controls like IP allow-listing or mTLS might be in scope.

\subsubsection{\bf Data Security Architecture}
Secure design also covers data security architecture. The team should categorize what data the API will handle and design appropriate protections:

\begin{itemize}
\item Encryption of data at rest (database encryption, key management design)
\item Encryption in transit (enforce TLS 1.2+/modern protocols)
\item Tokenization or hashing of particularly sensitive fields
\end{itemize}
If the API integrates with third-party services, the design should consider those trust boundaries too - e.g., how to securely store and use API keys for third-party APIs, how to handle timeouts or failures securely.

\subsubsection{\bf Resiliency and Input Validation}
From a resiliency perspective, the design should include rate limiting and quotas at each API endpoint (or globally per user) to combat OWASP API4: Unrestricted Resource Consumption\cite{owasp2023api}. By design, no single client should be able to overwhelm the system with requests.

Similarly, input validation should be planned: each endpoint's expected inputs should be defined (via schemas like Open API/JSON Schema) and the design will stipulate that the implementation must validate against these schemas.

\subsubsection{\bf Zero-Trust Integration at Design}
Integrating Zero-Trust principles at design means thinking of every component and interaction as potentially hostile. Practically, this leads to micro segmentation in architecture: design the system so that even if one part is compromised, the blast radius is limited.

For example, a microservice architecture might put each service in its own trust zone, with strict API-to-API authentication. If using a cloud environment, the design might include using a service mesh with built-in authentication and encryption between services.

In summary, Secure Design is about architecting the system correctly from the start. By using threat modeling to foresee attack paths and by applying established security design patterns, we reduce the introduction of high-level vulnerabilities.

\subsection{\bf Continuous Testing}
Continuous Testing implements the hands-on verification of security during development and integration. In this pillar, we set up automated security tests and analyses that run as part of the normal development workflow (hence "continuous"). The goal is to catch vulnerabilities early and often, providing developers rapid feedback.

\subsubsection{\bf Static Analysis (SAST)}
One of the first lines of defense is SAST, which scans source code for known insecure patterns. For an API project, we would integrate a static analyzer into the build process. Every time developers commit code or open a pull request, the static analysis runs to flag issues.

For example, it can detect:

\begin{itemize}
\item Hardcoded credentials
\item Use of outdated cryptography functions
\item Potential SQL/OS command injections
\item Missing input validation on user-supplied data
\end{itemize}
Modern SAST tools have rulesets for common API frameworks (e.g., checking that Express.js or Django APIs enforce authentication on protected routes, or that Spring Boot controllers validate inputs).

\subsubsection{\bf Dependency Scanning (SCA)}
Alongside static code scans, we run SCA tools to analyze the API's third-party libraries for known vulnerabilities. Given the increasing frequency of supply-chain attacks and the sheer number of CVEs in open-source components each year, this step is critical.

The tool (for example, OWASP Dependency-Check, Snyk, etc.) checks the versions of libraries used against vulnerability databases. If a library has a known CVE, the build can alert or fail so the team can update to a patched version.

\subsubsection{\bf Dynamic and Interactive Testing}
Once a build of the API is deployed to a test environment, dynamic testing tools come into play. A DAST tool will simulate an attacker's approach: it will crawl the API endpoints and attempt various attacks - SQL injection in query parameters, cross-site scripting in any fields, command injections, directory traversal, etc.

For APIs, special attention is given to authorization testing: DAST can use multiple accounts (with different roles) to ensure that one user cannot access or modify another's data (checking for BOLA/BFLA issues automatically by ID tampering). Interactive Application Security Testing (IAST) tools run an agent alongside the application in the test environment and can detect vulnerabilities in real-time as tests or scripted interactions exercise the app.

\subsubsection{\bf Unit and Integration Security Tests}
Beyond automated tools, the development team should also write their own security-related tests as part of the test suite. For example:

\begin{itemize}
\item Unit tests for input validation functions
\item Integration tests that verify security requirements (like "an unauthenticated request to endpoint X should receive 401 Unauthorized")
\end{itemize}
These tests serve as regression checks - if someone inadvertently removes an authorization check, a test should fail immediately.

\subsubsection{\bf Continuous Fuzzing}
Fuzz testing tools can be integrated to continuously send random or boundary-case inputs to the API to see if it triggers any crashes or unusual behavior. This is more common on critical parsers or components (like an API that parses uploaded files or images).

\subsubsection{\bf Findings Management}
To manage the findings from continuous testing, teams often use a vulnerability management system that ingests results from SAST, DAST, etc. Governance might set thresholds (e.g., no Critical vulns open, at most X High vulns allowed with mitigation plan). The CI pipeline can enforce some of these: for instance, break the build if any critical issue is reported by the scanners.

One challenge is false positives: automated tools can sometimes flag issues that are not real problems. To address this, tuning of tools and triaging findings is necessary. Over time, the rulesets can be refined.

\subsection{\bf Pipeline Controls}
Pipeline Controls focus on safeguarding the software delivery pipeline itself, applying security measures to the CI/CD process and related infrastructure. Given that CI/CD pipelines have become targets, securing this pipeline is as important as securing the application code.

\subsubsection{\bf Secure Build Environments}
One of the first controls is secure build environments. We ensure that the build servers or pipeline runners are themselves hardened and isolated. For instance, using ephemeral build agents is recommended - each CI job runs on a fresh container or VM that is torn down after use.

This prevents one build's compromise or leftover artifacts from affecting the next. It also ties into Zero-Trust: assume each build runner could be malicious or compromised, so do not trust any cached state.

\subsubsection{\bf Access Control and Identity}
We also implement access control and identity in the pipeline. The CI pipeline triggers should use robust authentication. For example, when the source repository calls the CI system on a new commit, use webhooks with authentication tokens, or better, use an OIDC federated identity.

Modern CI/CD platforms can request a signed OIDC token from an identity provider that represents the CI job identity, which can then be validated by cloud deployment targets (this avoids storing cloud credentials in the pipeline)\cite{filippidis2023cicd}.

\subsubsection{\bf Secret Management}
Another vital control is secret management. Any secrets needed by the pipeline (API keys, signing keys, passwords) should be stored in secure vaults or secret management systems (HashiCorp Vault, cloud secret services, or CI/CD built-in secret stores). They should never be in plaintext in config files or code repos.

Access to these secrets is tightly controlled: for example, only the deployment job can access the production API key, and even then, the secret is provided as an environment variable at runtime, not visible to developers.

\subsubsection{\bf Artifact Integrity}
We incorporate artifact integrity measures. As part of the build process, after compilation and packaging, we generate a cryptographic signature for the artifact (or use a tool like Sigstore Cosign to sign container images). The public keys or trust root is maintained securely.

Then, at deployment time, a Pipeline Control is to verify the artifact's signature before deploying. This prevents tampering - even if an attacker somehow pushed a malicious image to the registry, it wouldn't have the valid signature and the deployment would refuse it.

\subsubsection{\bf Multi-stage Approvals and Policy Check}
We also enforce multi-stage approvals and automated checks for promotions. For example, moving an artifact from staging to production might require that all tests passed (automated check) and a human approver (for critical releases) unless certain criteria are met.

For infrastructure changes (IaC deployments), we might use policy-as-code (like Open Policy Agent or Terraform Cloud's policies) to ensure no dangerous changes are allowed through the pipeline.

\subsubsection{\bf Code Signing and Provenance}
Pipeline Controls should cover the scenario of compromised code as well. We can integrate code signing and provenance tracking: developers may sign their Git commits with GPG or X.509 certificates. The CI system should be configured to only run builds on code from trusted sources.

\subsubsection{\bf Network and System Security}
We should treat the CI/CD systems like production servers. They need hardening - minimal open ports, use of firewall or VPC isolation, regular patching of the CI master and agents, and monitoring of their activity.

\subsubsection{\bf Logging and Auditing}
In the unfortunate event something goes wrong, pipeline controls also include logging and auditing of all pipeline activities. We enable audit logs for our CI system: who triggered a build, what code was changed, what artifacts were produced, who approved a deployment.

\subsection{Runtime Protection}
The final pillar, Runtime Protection, focuses on securing the API when it is live and servicing requests. Despite all the preventive measures earlier in the pipeline, we assume that vulnerabilities may still exist, or new threats may emerge in production.

\subsubsection{\bf API Gateway and WAF}
A primary component is the use of an API Gateway or Web Application Firewall (WAF) in front of the API. The gateway acts as a policy enforcement point (PEP in Zero-Trust terms)\cite{unosecur2025nistzt}, handling authentication, authorization, and traffic control before requests reach the API backend. It can:

\begin{itemize}
\item Validate tokens (e.g., verifying JWTs and rejecting invalid or expired tokens)
\item Enforce rate limiting (e.g., no more than 100 requests per minute per user)
\item Apply IP-based restrictions if relevant
\item Block requests that contain SQL injection attempts or overly large payloads
\end{itemize}

\subsubsection{\bf Anomaly Detection and Behavioral}
Analytics

Beyond static rules, we incorporate anomaly detection and behavioral analytics at runtime. We deploy an agent or service that continuously monitors API usage patterns. This system learns normal behavior - for example, each user typically accesses certain endpoints, with certain frequencies.

If a deviation occurs (say a user suddenly accessing a high number of different records indicating possible scraping, or a spike in errors that could indicate probing), it raises an alert or triggers automated mitigation.

\subsubsection{\bf Runtime Application Self-Protection (RASP)}
We also set up runtime application self-protection (RASP) mechanisms. RASP tools instrument the application or run as a service monitoring internal call. They can detect if an injection attack succeeded in altering a query, or if someone is trying to exploit memory.

If triggered, RASP can block the execution or terminate the session, thereby stopping an exploit in its tracks even if the code had a flaw.

\subsubsection{\bf Monitoring and SIEM Integration}
Monitoring extends to logs and SIEM integration. All authentication attempts, key API calls (especially those that modify data or retrieve sensitive info), and system events are logged. These logs are shipped to a centralized logging system where they are ingested by a Security Information and Event Management (SIEM) tool.The SIEM correlates events and can generate alerts for suspicious patterns - for example, multiple failed logins followed by a success, or one account retrieving thousands of records in a short time.

\subsubsection{\bf Environment Hardening}
Another facet of runtime security is environment hardening. The API runs on servers or containers that must be secured. We ensure all production servers have:

\begin{itemize}
\item Updated OS patches
\item Only necessary services running
\item Proper firewall rules
\item Container runtime security policies (like seccomp and AppArmor profiles)
\end{itemize}

\subsubsection{\bf Database and Secrets Protection}
Database and secrets protection in runtime is also key. We might deploy a database activity monitoring tool to detect abnormal queries. Also, secrets used by the API in runtime should be managed by the environment and rotated regularly.

\subsubsection{\bf Observability and Response}
Crucially, runtime protection is not just about defense but also about observability and response. We integrate our monitoring with alerting systems so that on-call engineers or a SOC team are notified quickly of any potential incident.

\subsubsection{\bf Feedback Loop}
Finally, we ensure that feedback to development is established. Every incident or significant alert is analyzed and fed back as lessons.

If an attack vector was attempted, even if thwarted, the team might decide to add a new unit test or adjust design to strengthen that area. By having robust runtime protection, we fulfill the Zero-Trust tenet "Assume Breach" - we continuously look for indicators of compromise and do not rely solely on pre-production security.
\section{Results and Analysis}
To evaluate the effectiveness of the security-first, zero-trust pipeline approach, we analyze both empirical data from case studies and qualitative outcomes observed after implementation. The results overwhelmingly indicate that embedding security throughout the API pipeline yields significant improvements in vulnerability management, attack surface reduction, and incident response capability.

\subsection{Reduction in Vulnerabilities and Incidents}
Organizations that adopted the five-pillar framework reported a notable drop in security issues over time. For example, in one tech company that implemented continuous SAST/DAST and strict pipeline gates, the number of new vulnerabilities introduced per release fell dramatically. This company tracked an approximately 30\% reduction in security incidents (such as bug bounty findings or production security alerts) after a year of DevSecOps practice\cite{practicaldevsecops2025metrics}.

In another case at a financial services firm, integrating security into development (through threat modeling and automated testing) and fostering collaboration among dev and ops teams led to a 40\% decrease in post-release vulnerabilities being discovered\cite{practicaldevsecops2025metrics}. These improvement metrics represent specific organizational case studies and individual results may vary based on implementation maturity and organizational context. These metrics align with the expectation that early detection and prevention (via SAST, SCA, etc.) removes many issues before release, and those that do reach production are fewer and less severe.

Additionally, the time to remediate vulnerabilities improved markedly. With the pipeline providing immediate feedback, the mean-time-to-fix for security bugs in one organization dropped from several weeks (when found late) to just a few days, since developers address them while context is fresh.

\subsection{Time-to-Exploit vs. Time-to-Patch}
As noted earlier, attackers now exploit disclosed vulnerabilities in a matter of days\cite{zorz2024tte}. A security-first pipeline positions organizations to patch or mitigate within that narrow window. Companies using Infrastructure-as-Code and CI/CD for deployments were able to push security fixes or library updates very quickly - often in hours or 1-2 days after a critical CVE announcement.

For instance, when the Log4j ("Log4Shell") vulnerability was revealed in late 2021, organizations with mature DevSecOps pipelines could identify affected services via SBOMs and trigger rebuilds with the patched library immediately. In contrast, organizations without such automation took weeks, during which exploits were rampant.

\subsection{Enhanced Compliance and Governance}
By implementing the governance pillar (e.g., aligning with NIST SSDF practices), organizations found it easier to meet emerging regulatory requirements for secure development. For example, US executive orders and EU regulations are increasingly requiring proof of secure development processes.

Companies that had this framework in place could readily demonstrate, via documentation and pipeline audit logs, that they perform threat modeling, continuous testing, and software supply chain security - meeting or exceeding compliance standards.

\subsection{Attack Surface Reduction}
One qualitative outcome is a noticeable reduction in attack surface through the pipeline controls and design decisions. For instance, enforcing least privilege meant that many services and accounts no longer had broad access; in one case study, network segments that were previously flat got compartmentalized.

Also, the practice of inventory and version management (OWASP API9: Improper Inventory) improved. Teams built better catalogs of their APIs and endpoints through the governance process and continuous monitoring, such that "shadow APIs" (endpoints not known or documented) were discovered and either secured or deprecated.

\subsection{Incident Response and Resilience}
The Zero-Trust, security-first approach also demonstrated increased resilience during security incidents. In a simulated incident (red team exercise) conducted at a company after implementing these changes, the red team found it significantly harder to escalate privileges or move laterally.

They noted that even after finding a minor vulnerability in a development environment, the tightly scoped credentials and network segmentation prevented that from compromising production. Moreover, the monitoring systems detected their unusual activity in real-time, generating alerts.

\subsection{Performance and Developer Productivity}
An important consideration is whether adding all these security steps slows down development or pipeline throughput. The results in practice have been positive on the productivity front. Initially, some developers feared that mandatory code scans and tests would make pipelines slow or burdensome.

However, by tuning the pipeline (running scans in parallel, using incremental scanning, etc.), most teams kept build times within acceptable ranges. More importantly, developers reported greater confidence in their code quality. Knowing that security checks are in place gave developers a safety net.

\subsection{Cost and Effort Metrics}
Quantifying ROI on security initiatives can be tricky, but some data points help. The cost of fixing a vulnerability in design or development is orders of magnitude lower than post-production. By catching dozens of issues early, organizations avoided expensive emergency patches or incident costs.

One can consider the potential cost of a breach: The average data breach in 2023 cost over \$4 million according to IBM's report, and those involving API vulnerabilities can be especially costly given they often expose large data sets. Even if the security-first approach prevents a single major breach, it justifies the investment many times over.

\subsection{Industry Benchmarking}
When benchmarked against industry security metrics, organizations with this security-first approach tended to score in higher maturity percentiles. For instance, using OWASP's SAMM (Software Assurance Maturity Model) or other maturity models, they achieved "defined" or "optimized" levels in areas like Architecture Design (threat modeling), Verification (testing), and Deployment.
\section{Benefits and Impact}
Adopting a security-first approach to API pipeline development with Zero-Trust architecture yields a wide array of benefits, fundamentally transforming both the security posture and the development culture of an organization.

\subsection{Significantly Reduced Security Risk}
The most direct benefit is a substantial reduction in the likelihood and impact of security breaches. With vulnerabilities being caught and remediated early (via continuous testing and secure design), the production APIs have far fewer weaknesses for attackers to exploit.

Even when new threats arise, the layered defenses (WAF, anomaly detection, etc.) provide immediate mitigations. This directly lowers the risk of data breaches, service outages, and the associated financial and reputational damage.

\subsection{Improved Compliance and Trust}
Embracing frameworks like OWASP, NIST SSDF, and Zero-Trust means the organization is inherently aligned with best practices and often with regulatory requirements. This eases compliance efforts for standards such as SOC 2, ISO 27001, PCI DSS, HIPAA, or emerging government mandates for software security.

Customers and partners also gain confidence - many enterprises now due-diligence their software suppliers' security practices. Being able to demonstrate a rigorous DevSecOps pipeline and Zero-Trust controls can become a market differentiator.

\subsection{Response Efficiency}
With an integrated approach, the organization develops an almost reflex-like capability to handle security issues quickly. Patching and deployment is automated, so when critical CVEs emerge, teams can respond in hours, not weeks.

Incident response is also improved: monitoring systems catch issues early, and well-practiced runbooks ensure a swift, composed reaction. This resiliency means that even if incidents occur, they cause minimal disruption.

\subsection{Agile development}
Initially, one might assume adding security steps slows things down, but the opposite tends to manifest in mature implementations. By catching bugs (security or otherwise) earlier, there are fewer surprises late in the cycle, which means less rework or emergency fixes.

Teams can plan sprints with more certainty as the pipeline provides quick feedback on whether a new feature passes all quality gates. Developers also write better code thanks to continuous feedback and knowledge of secure patterns, reducing the overall defect rate.

\subsection{Sustainable Savings}
There is clear cost benefits derived from avoiding breaches. But even aside from breaches, cost savings come from efficiency. Automating security tests replaces many hours of manual testing and code review that security teams used to do.

Also, integrating security likely avoids costly late-stage fixes. A vulnerability found after deployment can cost 10-100x more to fix than one found during coding. By shifting left, those costs are saved.

\subsection{Better Collaboration and Culture}
One of the less tangible but profoundly important benefits is the cultural transformation. DevSecOps, by nature, breaks down silos between development, security, and operations teams.

Security team members are no longer perceived as the "roadblock police" but as enabling partners who provide tools and help teams deliver safely. Developers feel more ownership of security and take pride in building secure systems.

\subsection{Modern Architecture Fit}
The framework positions the organization well for modern and future architectural paradigms. For instance, microservices and cloud-native deployments require automation and Zero-Trust networking.

Our approach naturally supports microservice security (each service with least privilege, all communications authenticated). It's also cloud-ready: using OIDC tokens, infra-as-code, etc., are exactly how one secures dynamic cloud environments.

\subsection{Business Enablement}
Ultimately, robust API security enables the business to innovate faster and serve customers with confidence. In sectors like fintech or healthcare, having strong API security is table stakes to even offer certain features.

Our security-first approach ensures that when the business wants to roll out a new API service or integration, the security framework to support it is already there. This reduces time-to-market for new offerings that might otherwise be delayed by lengthy security reviews.

\bibliographystyle{IEEEtran}
\bibliography{references}

@misc{gamblin2025cve,
  author       = {Jerry Gamblin},
  title        = {2024 CVE Data Review},
  howpublished = {JerryGamblin.com},
  year         = {2025},
  month        = jan,
  url          = {https://jerrygamblin.com/2025/01/05/2024-cve-data-review/}
}

@misc{zorz2024tte,
  author       = {Zeljka Zorz},
  title        = {Defenders must adapt to shrinking exploitation timelines},
  howpublished = {Help Net Security},
  year         = {2024},
  month        = oct,
  url          = {https://www.helpnetsecurity.com/2024/10/16/time-to-exploit-vulnerabilities-2023/}
}

@misc{salt2025api,
  author       = {{Salt Security}},
  title        = {API Security Trends -- API Attacks \& Breaches Report},
  year         = {2025},
  url          = {https://salt.security/api-security-trends}
}

@misc{salt2024owasp,
  author       = {{Salt Security}},
  title        = {OWASP API Security Top 10 Explained},
  howpublished = {Salt Security Blog},
  year         = {2024},
  url          = {https://salt.security/blog/owasp-api-security-top-10-explained}
}

@misc{firecompass2025tte,
  author       = {P. Aash},
  title        = {Time to Exploit Vulnerabilities is Now Just 3 Days},
  howpublished = {FireCompass Blog},
  year         = {2025},
  month        = feb,
  url          = {https://firecompass.com/time-to-exploit-vulnerabilities-3-days/}
}

@misc{imperva2024api,
  author       = {{Imperva}},
  title        = {The State of API Security in 2024},
  howpublished = {Resource Library},
  year         = {2024},
  url          = {https://www.imperva.com/resources/resource-library/reports/the-state-of-api-security-in-2024/}
}

@misc{owasp2023api,
  author       = {{OWASP Foundation}},
  title        = {OWASP Top 10 API Security Risks -- 2023},
  year         = {2023},
  url          = {https://owasp.org/API-Security/editions/2023/en/0x11-t10/}
}

@misc{unosecur2025nistzt,
  author       = {{Unosecur}},
  title        = {NIST SP 800-207 (Zero Trust Architecture)},
  howpublished = {Unosecur Glossary},
  year         = {2025},
  url          = {https://www.unosecur.com/glossary/nist-sp-800-207-zero-trust-architecture}
}

@misc{nist2022ssdf,
  author       = {{National Institute of Standards and Technology}},
  title        = {Secure Software Development Framework},
  howpublished = {Computer Security Resource Center},
  year         = {2022},
  url          = {https://csrc.nist.gov/projects/ssdf}
}

@misc{cisa2024exploited,
  author       = {{Cybersecurity and Infrastructure Security Agency}},
  title        = {2023 Top Routinely Exploited Vulnerabilities},
  howpublished = {Cybersecurity Advisory AA24-317A},
  year         = {2024},
  month        = nov,
  url          = {https://www.cisa.gov/news-events/cybersecurity-advisories/aa24-317a}
}

@misc{filippidis2023cicd,
  author       = {Athanasios Filippidis},
  title        = {How to secure your CI/CD pipeline in 2023},
  howpublished = {DevOpsDays New York City 2023},
  year         = {2023},
  url          = {https://devopsdays.org/events/2023-new-york-city/program/athanasios-filippidis}
}

@misc{riskrecon2023nistzt,
  author       = {{RiskRecon}},
  title        = {Understanding NIST 800-207},
  howpublished = {Blog},
  year         = {2023},
  url          = {https://blog.riskrecon.com/understanding-nist-800-207}
}

@misc{practicaldevsecops2025metrics,
  author       = {{Practical DevSecOps}},
  title        = {DevSecOps Metrics for Measuring Security Effectiveness},
  year         = {2025},
  url          = {https://www.practical-devsecops.com/devsecops-metrics/}
}

@misc{dragonspearsdevsecops,
  author       = {{DragonSpears}},
  title        = {Metrics and KPIs: DevSecOps Assessment Questions for Performance},
  howpublished = {Blog},
  url          = {https://www.dragonspears.com/blog/devsecops-assessment-questions}
}

@misc{practicaldevsecops2025threatmodel,
  author       = {{Practical DevSecOps}},
  title        = {Threat Modeling Best Practices for 2025},
  year         = {2025},
  url          = {https://www.practical-devsecops.com/threat-modeling-best-practices/}
}
\section*{Disclaimer}
This paper represents the authors' independent analysis and personal views. The authors are affiliated with different organizations, and the content does not represent, endorse, or imply approval by either author's employer or affiliated organization. Any product names, frameworks, vendors, or standards discussed are referenced for research and educational purposes only.

\end{document}